\documentclass[epj]{svjour}

\usepackage{graphicx}
\usepackage{hyperref}
\usepackage{braket}
\usepackage{amsmath,amssymb}
\usepackage{csquotes}
\usepackage{multirow}
\usepackage{floatrow}
\usepackage{rotating}
\usepackage{xcolor}
\usepackage{upgreek}
\usepackage[flushleft]{threeparttable}
\usepackage{enumitem}

\begin{document}

\title{Dispersive Corrections in Elastic Electron-Nucleus Scattering: An Investigation in the Intermediate Energy Regime And their Impact on the Nuclear Matter}

\author{\textbf{The Jefferson Lab Hall A Collaboration}\thanks{\email{gueye@nscl.msu.edu}}}
\institute{Newport News, VA 23606, USA}

 
\abstract{
\mbox{}
Measurements of elastic electron scattering data within the past decade have highlighted two-photon exchange contributions as a necessary ingredient in theoretical calculations to precisely evaluate hydrogen elastic scattering cross sections. This correction can modify the cross section at the few percent level. In contrast, dispersive effects can cause significantly larger changes from the Born approximation. The purpose of this experiment is to extract the carbon-12 elastic cross section around the first diffraction minimum, where the Born term contributions to the cross section are small to maximize the sensitivity to dispersive effects. The analysis uses the LEDEX data from the high resolution Jefferson Lab Hall A spectrometers to extract the cross sections near the first diffraction minimum of $^{12}$C at beam energies of 362~MeV and 685~MeV. The results are in very good agreement with previous world data, although with less precision. The average deviation from a static nuclear charge distribution expected from linear and quadratic fits indicate a 30.6\% contribution of dispersive effects to the cross section at 1~GeV.  The magnitude of the dispersive effects near the first diffraction minimum of $^{12}$C has been confirmed to be large with a strong energy dependence and could account for a large fraction of the magnitude for the observed quenching of the longitudinal nuclear response. These effects could also be important for nuclei radii extracted from parity-violating asymmetries measured near a diffraction minimum.
}

\authorrunning{P.~Gu\`eye}

\titlerunning{Investigation of Dispersive Corrections to the Born Approximation \ldots}
\maketitle

\section{\label{sec:intro}Introduction\protect}

During the 80s and 90s, higher order corrections to the first Born approximation were extensively  studied through dedicated elastic and quasi-elastic scattering experiments using unpolarized  electron and positron beams (see~\cite{Cardman:1980dja,Offermann:1986en,Breton:1991fe,Offermann:1991ft,Gueye:1998zz,Gueye:1999mm}  and references therein), following the seminal paper from~\cite{Sick:1970ma}. These effects scale as $S_{HOB} = V_C/E_e$ where $S_{HOB}$ is the scaling factor to account for higher order corrections to the Born approximation, $V_C$ is the Coulomb potential of the target nucleus and $E_e$ is the incident energy of the lepton probe~{\color{black}\cite{Gueye:1999mm}}.  Incidentally, they are expected to be small in the medium to intermediate energy regime, and have been neglected in the analysis of GeV energy data: $V_C$ reaches a maximum of about 26~MeV for $^{208}$Pb with a corresponding value of $S_{HOB} =  0.52\%$ for a 5~GeV beam.

\begin{figure}[!htbp]
	\centering
		\includegraphics[width=1.\textwidth]{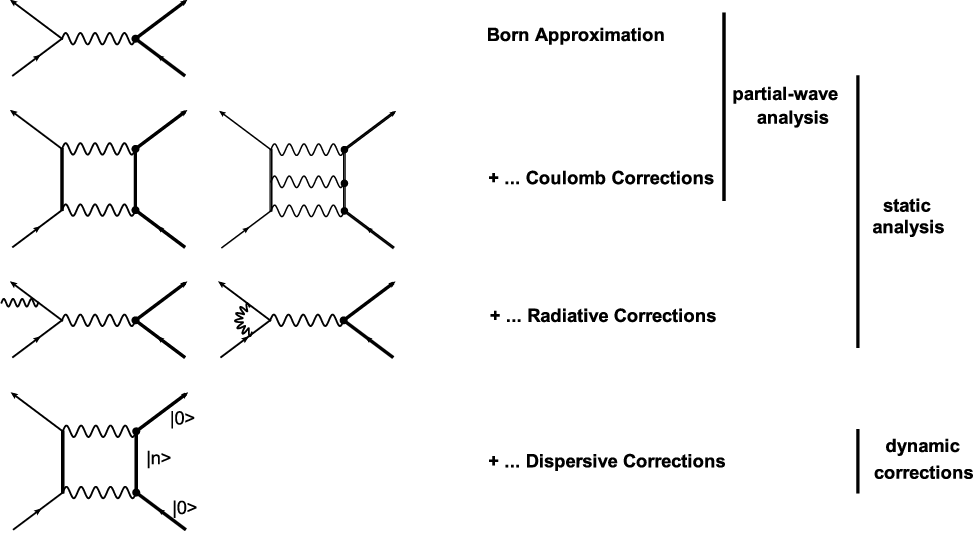}
	\caption{\label{fig:HOB}\protect  High-order corrections to the one-photon exchange Born approximation in electron/positron-nucleus scattering.}
\end{figure}

In the 1$^{\rm st}$ order approximation, the scattering cross section is evaluated using plane wave functions for the incoming and outgoing electrons. This approach is also known as the Plane Wave Born approximation (PWBA) or simply the Born Approximation (Fig.~\ref{fig:HOB}). Coulomb corrections originate from the Coulomb field of the target nucleus that causes an acceleration (deceleration) of the incoming (outgoing) electrons and a Coulomb distortion of the plane waves: these effects are treated within a Distorted Wave Born Approximation (DWBA) analysis for inelastic scattering or elastic/quasi-elastic scattering on heavy nuclei~\cite{Gueye:1999mm}. Two other corrections are required to properly evaluate the scattering cross section: radiative corrections due to energy loss processes and dispersive effects due to virtual excitations of the nucleus at the moment of the interaction (Fig.~\ref{fig:HOB}).

Within the last decade, a renewed interest arose from a discrepancy between unpolarized and polarized elastic scattering data on the measurement of the proton form factor ratio $\mu G^p_E/G^p_M$ which can be attributed to the contribution of two-photon exchanges~\cite{Guichon:2003qm,Blunden:2003sp,Rekalo:2004qa,Chen:2004tw,Afanasev:2004hp,Arrington:2004ae,Blunden:2005ew,Blunden:2017nby}. These effects have been investigated with a series of dedicated experiments~\cite{Adikaram:2014ykv,Rachek:2014fam,Henderson:2016dea,Rimal:2016toz} (also see reviews~\cite{Carlson:2007sp,Arrington:2011dn,Afanasev:2017gsk} and references therein), including their impact on  the measurement of form factors for nucleons and light ($A\leq 3$) nuclei. They include both Coulomb corrections~\cite{Gueye:1999mm,Aste:2005wc}, excited intermediate states and treatment of the off-shell nucleons through dispersion relations as a function of the 4-momentum transfer. 

Coulomb corrections have historically been labeled as \textit{static} corrections to the Born approximation as depicted in Fig.~\ref{fig:HOB}. While these effects contribute to a few percents~\cite{Gueye:1999mm,Carlson:2007sp,Arrington:2011dn,Aste:2005wc}, \textit{dynamic} corrections known as dispersive effects are emphasized in the diffraction minima, where the first-order (Born approximation) cross section has a zero, and can contribute up to {\color{black}18\% in the first} diffraction minimum of $^{12}$C at 690~MeV \cite{Offermann:1991ft,Gueye:1998zz}.

The electromagnetic nuclear elastic cross section for electrons can be expressed as:
\begin{equation}\label{eq:xsec_theo}
\frac{d\sigma}{d\Omega} = \Big( \frac{d\sigma}{d\Omega} \Big)_{Mott} \mid F(q^2)\mid ^2
\end{equation}
\noindent
where $\Big ( \frac{d\sigma}{d\Omega} \Big )_{Mott}$ is the Mott cross section corresponding to the scattering on a point-like nuclear target, $F(q^2)$ represents the form factor and $q^2 = -Q^2$ is the 4-momentum transfer. 

Theoretical calculations for dispersive effects in elastic electron scattering for p-shell, spin-0 targets such as $^{12}$C were performed in the mid-70s by Friar and Rosen~\cite{Friar:1974bn}. They used a harmonic oscillator model and only the longitudinal (Coulomb) component to calculate the scattering amplitude within the {\color{black}PBWA} approximation; the transverse component was neglected. The matrix element in the center-of-mass frame -- considering only the contribution from the dominant two photon exchange diagrams -- can be written as:
\begin{equation}\label{eq:M_disp}
{\cal{M}}_{disp} = \sum _{n \neq 0} \int \frac{d^3\vec{p}}{\vec{q}_1^2\vec{q}_2^2}
			\frac{\braket{0 | \rho(\vec{q}_2) | n}\braket{n | \rho(\vec{q}_1) | 0}}{p^2-p^2_n-i\varepsilon}a(p_n)
\end{equation}
\noindent
with:
\begin{equation}\label{eq:M_disp_parameters}
\left\{
\begin{array}{lll}
a(p_n)	& = & E_e p_n [1 + \cos\theta] + \vec{p}\cdot (\vec{p_{e}} + \vec{p}_{e'})	\\
p_n		& = & E_e - \omega_n - \frac{p^2 - E_e^2}{2M_p}					\\
p		& = & p_e - p_{e'}
\end{array}
\right.
\end{equation}
\noindent
where: $p_e = (E_e, \vec{p}_e)$ and $p_{e'} = (E_{e'}, \vec{p}_{e'})$ the 4-momentum of the incoming and outgoing electrons, respectively, and $\vec{q}_{1,2}$ the 3-momenta of the two photons exchanged. $\theta$ is the angle between the incoming and outgoing electrons. $\rho(\vec{q}_1)$ and $\rho(\vec{q}_2)$ are the charge operators associated with the two virtual photons, respectively, and using the notation of~\cite{Friar:1974bn} with $\hat{e}_i(\vec{q})$ the charge distribution (operator in the isospin space) of the $\rm i^{th}$ nucleon, gives:

\begin{equation}\label{eq:rho_q1q2}
\mbox{\large\(
\left\{
\begin{array}{lll}
\rho(\vec{q})	& = & \sum _{i=1}^{A}\hat{e_i}(\vec{q})e^{i\vec{q}\cdot \vec{x'}_i} 	\\
			&  & \\
\hat{e}(\vec{q})	& = & \int \hat{e}(\vec{x})e^{i\vec{q}\cdot \vec{x}}d^3\vec{x}
\end{array}
\right.
\)}
\end{equation}

In their calculation, Friar and Rosen~\cite{Friar:1974bn} also considered that all nuclear excitation states $\ket{n}$ have the same mean excitation energy $\omega$, allowing to apply the closure relation: 
$\sum \ket{n}\bra{n} = 1$. Including the elastic scattering and dispersion corrections leads to:
\begin{equation}\label{eq:M_disp_FF}
{\cal{M}}_{elast + disp} = (\alpha Z)F(q^2) + (\alpha Z)^2 G(q^2)
\end{equation}
\noindent
with $G(q^2)$ arising from two-photon exchange diagrams (including cross-diagram, seagull \ldots). Hence:
\begin{equation}\label{eq:M_disp_sq}
\left.
\begin{array}{lll}
|{\cal{M}}_{elast + disp}|^2
						& = & (\alpha Z)^2 \big [ F(q^2) \big ]^2	\\
						&    & \\
						& + & 2(\alpha Z)^3 \big [ F(q^2) {\cal{R}}e\{G(q^2)\} \big ] \\
						&    & \\
			   			& + & (\alpha Z)^4 \big [ |{\cal{R}}e\{G(q^2)\} |^2 +  |{\cal{I}}m\{G(q^2)\}|^2 \big ]
\end{array}
\right.
\end{equation}

Therefore, the scattering amplitude is governed by $F(q^2)$ and the real part of $G(q^2)$ outside the minima of diffraction (where $F(q^2) \neq 0$). The imaginary part of $G(q^2)$ is most important in the minima of diffraction where {\color{black}the term $F(q^2)$ goes to zero}.

Experimentally, in order to extract the magnitude of the dispersive effects, the momentum transfer $q$ is modified to account for the Coulomb effects into an effective momentum transfer $q_{eff}$ {\color{black}(we refer the reader to~\cite{Gueye:1999mm,Aste:2005wc,Traini:2001kz} for the validity of this so-called Effective Momentum Approximation)}. The latter is obtained by modifying the incident $(E_e)$ and scattered $(E_{e'})$ energies of the incoming and outgoing electrons~\cite{Gueye:1999mm}:
\begin{equation}\label{eq:qeff}
q = 4E_eE_{e'}\sin ^2 (\theta/2 ) \rightarrow q_{eff} = 4E_{e,eff}E_{e',eff}\sin ^2 (\theta/2 )
\end{equation}
\noindent
with $E_{e,eff} = E_e\Big( 1 - \frac{|V_C|}{E_e} \Big)$ and $E_{e',eff} = E_{e'}\Big( 1 - \frac{|V_C|}{E_e} \Big)$. $|V_C|$ is the (magnitude of the) Coulomb potential of the target nucleus.

The corresponding experimentally measured cross section can then be compared to the theoretical cross section calculated using a static charge density~\cite{Offermann:1991ft}. This paper reports on a recent analysis of these effects in the first diffraction minimum of $^{12}$C at $q_{eff} \approx 1.84~{\rm fm}^{-1}$ performed in the experimental Hall A at Jefferson Lab~\cite{Ledex,Kabir:2015}.

\section{\label{sec:ledex}The LEDEX experimental setup\protect}

The Low Energy Deuteron EXperiment (LEDEX)~\cite{Ledex} was performed in two phases: first in late 2006  with a beam energy of 685 MeV and then in early 2007 with a beam energy of 362 MeV. They both used a 99.95 $\%$ pure $^{12}$C target with a density of 2.26 ${\rm g/cm^{3}}$ and a thickness of $0.083\pm 0.001~{\rm g/cm^2}$. The combined momentum transfer range was $0.4-3.0~{\rm fm^{-1}}$.

The two identical high-resolution spectrometers (HRS) \cite{Alcorn:2004sb} in Hall A were designed for nuclear-structure studies through the $(e,e^\prime p)$ reaction. Each contains three quadrupoles and a dipole magnet, all superconducting and cryogenically cooled, arranged in a QQDQ configuration. While the first quadrupoles focus the scattered particles, the dipole steers the charged particles in a $45^\circ$ upward angle, and the last quadrupole allows one to achieve the desired horizontal position and angular {\color{black}resolutions}. The HRS detector systems are located behind the latter to detect scattered electrons or electro-produced/recoiled hadrons. Each contains a pair of vertical drift chambers for tracking purpose~\cite{Fissum:2001st}, a set of scintillator planes,  a \v{C}erenkov detector~\cite{Iodice:1998ft}  and a {\color{black}two-layered} calorimeter for particle identification. During the LEDEX experiment, both spectrometers were tuned to detect elastically scattered electrons. The electrons which do not interact with the target are transported in a beam pipe and eventually  stopped in a beam dump located about 20~m downstream of the target.

The position of the left HRS (with respect to the incident beam direction) was changed  according to the kinematic settings while the right HRS was fixed at $24^\circ$ for calibration purposes.  The study of the optics for each of the HRS spectrometers was performed with tungsten sieve plates that were mounted in front of each spectrometer. These plates each have a 7 by 7 pattern of holes.  Two holes have a diameter of 4~mm while the remaining holes have a 2~mm diameter. The larger holes are placed asymmetrically so that their orientation in the image at the focal plane can be identified without any ambiguity.
Further details on this experimental setup can be found in~\cite{Lee:2009zzp}.

\begin{figure}[!htbp]
	\centering
		\includegraphics[width=0.8\textwidth]{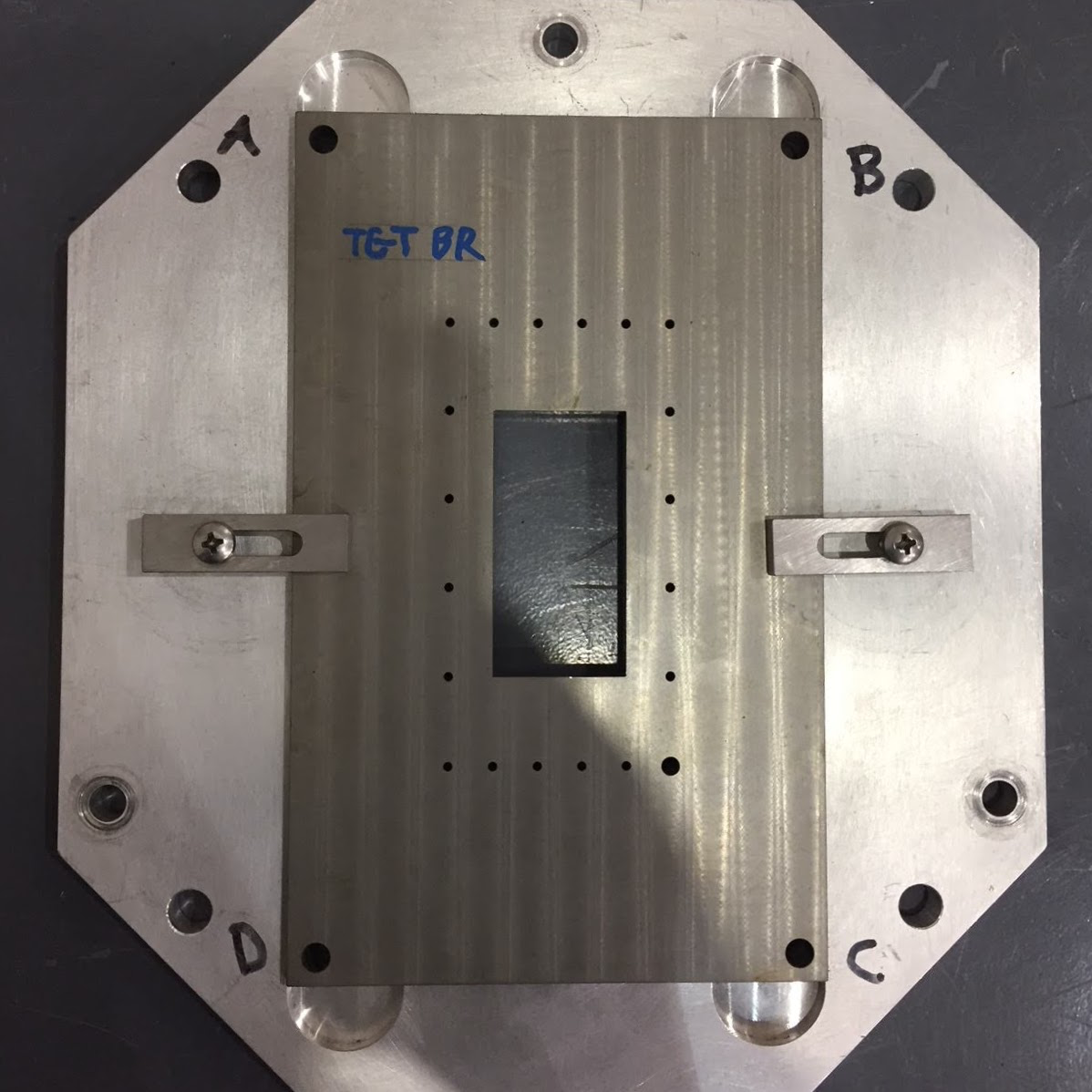}
		\caption{\label{fig:photo}\protect Photo of the tungsten (grey) 2~mrs collimator with its outer sieve holes that was used during the LEDEX experiment.  The outer aluminum frame mounted to the face of the HRS spectrometer with mounting bolts located at A,B,C and D. The tungsten plate could be removed if full HRS acceptance was desired without removing the outer aluminum frame. Sieve photo courtesy of Jessie Butler.}
\end{figure}

\begin{figure}[!htbp]
	\centering
		\includegraphics[width=0.9\textwidth]{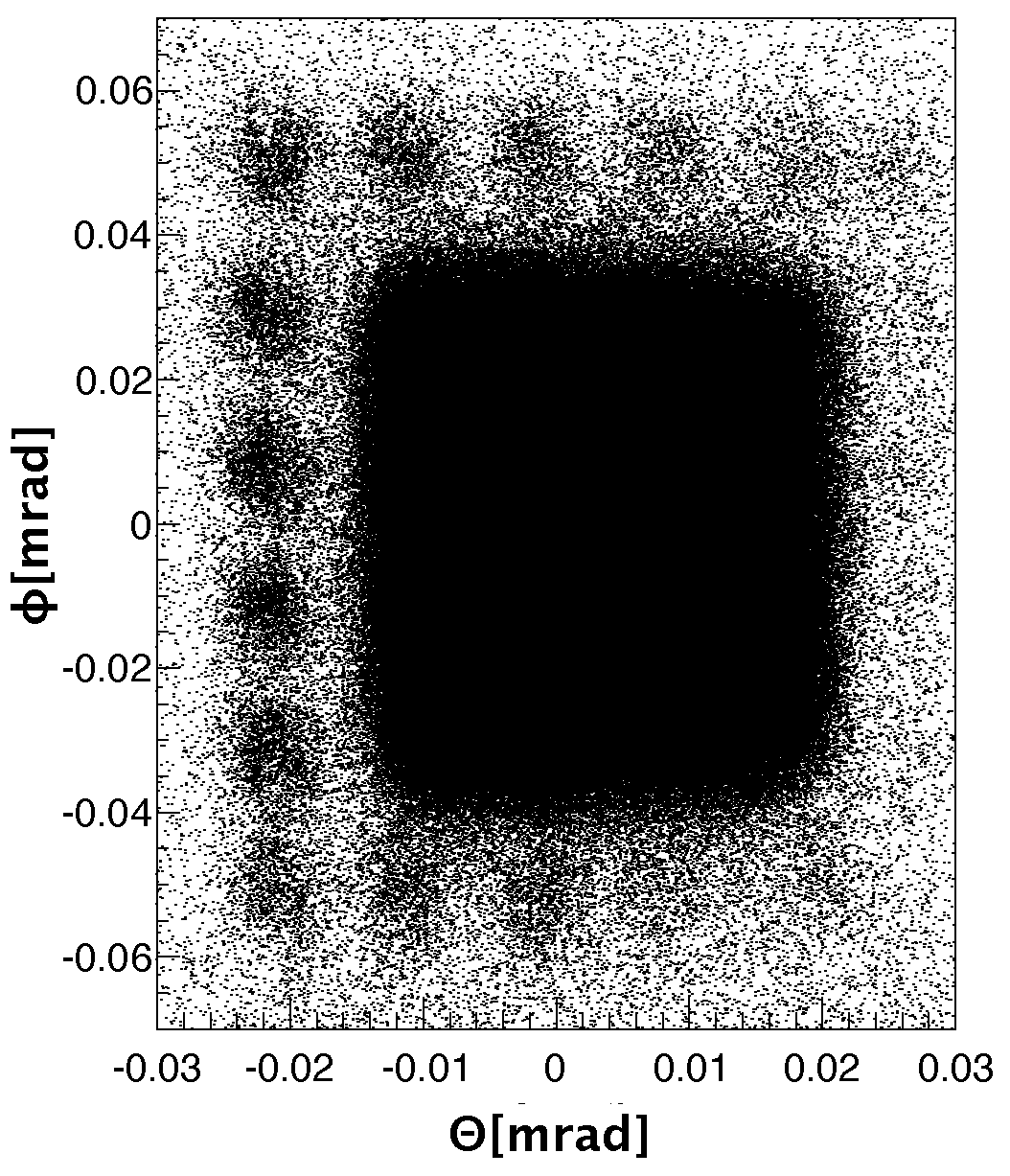}
		\caption{\label{fig:lhrs_collimator_data}\protect The experimentally reconstructed scattering ($\Uptheta$) and azimuthal ($\upphi$) spectrometer angles with the tungsten collimator installed.  The 2~msr opening is clearly visible. Due to the rapid decrease in the elastic cross section, only the small scattering angle sieve holes are visible.}
\end{figure}

For the elastic measurements, a 2~msr tungsten collimator was mounted to the face of the spectrometers: it has a $3\times 6~{\rm cm^2}$ rectangular hole at its center, nineteen 2~mm diameter pin holes symmetrically placed around it and one 4~mm diameter pin hole in the bottom corner of the central large opening as shown in Fig.~\ref{fig:photo}. The physical locations of these holes were surveyed before the start of the experiment. This redundant calibration check is performed to eliminate any ambiguity in the scattering angle (Fig.~\ref{fig:lhrs_collimator_data}): the 2D distribution of the spectrometer angles $\Uptheta$ (horizontal) and $\upphi$ (vertical) shows an asymmetric trapezoid reflecting the dependence of the cross section when going horizontally from -0.03~mrad (lower scattering angle) to 0.03~mrad (larger scattering angle).  

\section{\label{sec:data_anaysis}Data analysis\protect}

\noindent
The differential elastic scattering cross-sections were measured using Eq.~\eqref{eq:xsec_exp}:
\begin{equation}\label{eq:xsec_exp}
\frac{d\sigma}{d\Omega}=\frac{P_{S}\times N_{net}}{L\times t\times \Delta \Omega \times \Pi_i \epsilon_i}\times R
\end{equation}
\noindent where: $P_{S}$ is the pre-scale factor, $N_{net}$ is the net counts ({\color{black}found after applying necessary acceptance and particle identification cuts}), $L$ is the luminosity of the run, $t$ is the duration of the run, $\Delta \Omega$ is the solid angle, $\Pi_i \epsilon_i$ is the running (electronics, computer and cuts) efficiencies and $R$ is the radiative corrections factor. The luminosity for fixed target is calculated from $L= {\cal{F}}_e d_T l$, with ${\cal{F}}_e$ the incident particle flux, $d_T$ the density of the target, and $l$ the target thickness. 

Each gas \v{C}erenkov detector within the HRS spectrometers which allows for $\pi^-/e^-$ discrimination has a measured efficiency greater than 99.6\% for our experiment~\cite{Kabir:2015}: a {\color{black}pion} with a momentum of at least 4.8 GeV/c is required to produce a \v{C}erenkov light in this detector that is well above our maximum available beam energy of 0.686 GeV.

Only certain events were identified as ``good'' events: they consisted of events that have a single track, with one cluster per plane and a number of hits between 3-6 in addition to originating from the trigger level 3 (level 1) for the left (right) arm good track cuts on the vertical drift chambers. The tracking and triggering efficiencies were folded in the analysis when calculating the cross section.

Some ``good'' events were observed outside the physical acceptance of the spectrometer even within the calibrated data sets. These events were excluded using the geometrical cuts from the targets as well as the angular spectrometer acceptances~\cite{Kabir:2015}. The cuts were chosen to limit the data away from the edges of the acceptances where the distribution of these parameters varies rapidly. A further study of the ``white spectrum'' shows that the acceptance for both spectrometers is $\pm 3.9\%$, which is lower than the expected value of $\pm 4.5\%$. A tight cut of $\pm 3.9\%$ was applied on the momentum acceptance during the yield calculations.

The radiative corrections factor, $R$, cannot be evaluated experimentally: the MCEEP-Monte Carlo simulation code for ($e,e^\prime p$)~\cite{mceep} was used for that purpose. In MCEEP, the virtual photons are taken into account through a Schw\-inger term~\cite{Schwinger:1949zz}, found by the Penner calculation. The elastic radiative tail due to hard photons is approximated from the prescription by Borie and Dreschel~\cite{Borie:369B}, and Templon et al.~\cite{Templon:4607T} which is a corrected version of the original calculations from Mo and Tsai~\cite{Mo:1968cg}. MCEEP also accounts for the external radiation sources such as straggling, external Bremsstrahlung, energy losses  from multiple collisions with the atomic electrons etc. This simulation package was also used to calculate the phase space factors \cite{mceep}. Dead times (both electronic and computer) were found to be negligible for this experiment, and the tracking and triggering efficiencies found to be more than 99\%.

The maximum beam current achieved was $19.5~\mu A$ at 362~MeV and $23.4~\mu A$ at 685~MeV. Table~\ref{table:systematics} lists the primary sources of systematic uncertainties for the LEDEX experiment. Not listed is the uncertainty on the incident beam position of $\pm 200~\mu m$. Around the diffraction minima, the statistical uncertainty dominates translating to 7.70\% (statistical) and 3.50\% (systematic) at 362~MeV and 4.24\% (statistical) and 2.40\% (systematic) at 685~MeV. The situation is exactly the opposite outside the diffraction minima~\cite{Kabir:2015}. 

\begin{center}
\begin{table}[h]
\caption{Systematic uncertainties for the LEDEX experiment~\protect\cite{Kabir:2015}.}
\begin{tabular}{lcc}
\hline
Quantity 				& Normalization	& Random	\\
					& (\%)			& (\%)		\\
\hline
Beam Energy 			& 0.03			&  ---			\\
Beam Current 			& 0.50			&  ---			\\
Solid Angle 			& 1.00			& ---			\\
Target Composition 		& 0.05			& ---			\\
Target thickness		& 0.60			& ---			\\
Tracking Efficiency 		& --- 				& 1.00		\\
Radiation correction 		& 1.00			& ---			\\
Background Subtraction 	& ---				& 1.00		\\
\hline
\end{tabular}
\label{table:systematics}
\end{table}
\end{center}
\vspace{-0.75cm}

\begin{figure}[!htbp]
	\centering
		\includegraphics[width=1.\textwidth]{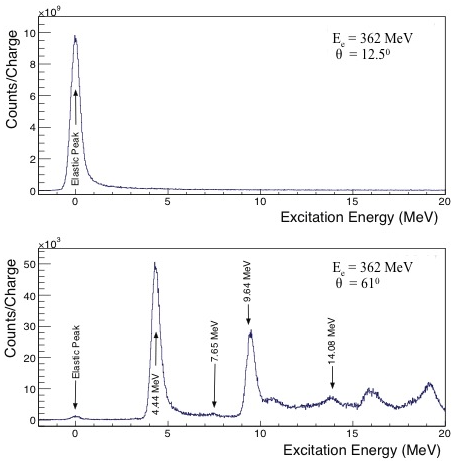}
		\caption{\label{fig:omega_dist_362MeV}\protect  The reconstructed excitation energy distributions at $E_e = 362$~MeV for $\theta = 12.5^\circ$ (top) and $\theta= 61^\circ$ (bottom) scattering angles.}
\end{figure}
\begin{figure}[!htbp]
	\centering
		\includegraphics[width=1.\textwidth]{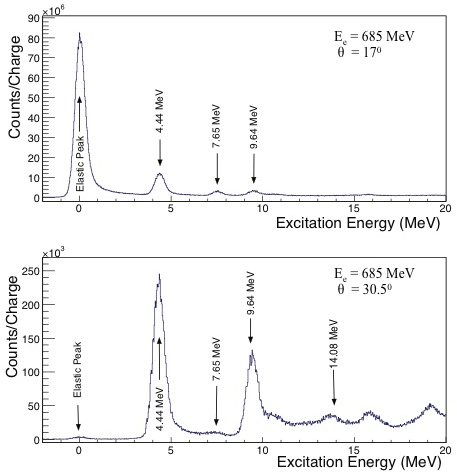}
		\caption{\label{fig:omega_dist_685MeV}\protect  The reconstructed excitation energy distributions at $E_e = 685$~MeV for $\theta = 17^\circ$ (top) and $\theta = 30.5^\circ$ (bottom) scattering angles.}
\end{figure}

Figures~\ref{fig:omega_dist_362MeV} and~\ref{fig:omega_dist_685MeV} show the reconstructed excitation energy distributions at 362~MeV and 685~MeV incident beam energies, respectively. The high resolution of the HRS spectrometers (0.05\%) allows to clearly identify the first four excited states of $^{12}$C for both energies: 4.44~MeV ($2^+$), 7.65~MeV ($0^+$), 9.64~MeV ($3^-$) and 14.08~MeV ($4^+$). This paper reports on the analysis of the elastic peak data.

\section{\label{sec:results}Results\protect}
\noindent
Table~\ref{table:qeff} lists the kinematics of the LEDEX experiment inside the first diffraction minimum of $^{12}$C that correspond to 4-momentum transfers $q$ of 1.85~${\rm fm}^{-1}$ and 1.82~${\rm fm}^{-1}$ ($q_{eff}$ of 1.82~${\rm fm}^{-1}$ and 1.81~${\rm fm}^{-1}$) for (362~MeV, $61^\circ$) and (685~MeV, $30.5^\circ$), respectively. The corresponding measured elastic cross sections are given in Table~\ref{table:xsec} and are found to be: $(3.26 \pm 0.28) \times 10^{-8}~{\rm fm}^2/{\rm sr}$ for 362~MeV and $(2.35 \pm 0.11) \times 10^{-7}~{\rm fm}^2/{\rm sr}$ for 685~MeV. They were compared to static cross sections calculated from a Fourier-Bessel (FB) parameterization extracted from the LEDEX data that is found to be almost identical to the one from Offermann et al.~\cite{Offermann:1991ft} and the agreement is within 0.1\%. A forthcoming paper on the Boron radius~\cite{Kabir:2015} discusses in more details the validity of this parameterization.
\begin{center}
\begin{table}[h]
\caption{The four-momentum transfer ($q$) and effective four-momentum transfer ($q_{eff}$) for the LEDEX experiment for each elastic kinematic setting calculated using \protect{Eq.~\eqref{eq:qeff}}.}
\begin{tabular}{c|c|c|c|c}
\hline
$E_e$	        & $\theta$  & $E_{e'}$	& $q$ 	        & $q_{eff}$ 	\\
(MeV)			& (Deg.)	& (MeV)		& (fm$^{-1}$)	& (fm$^{-1}$)	\\
\hline
362				& 12.5		& 361.72	& 0.40	    	& 0.39		    \\
362				& 61.0		& 356.06	& 1.85		    & 1.82		    \\
\hline
685				& 17.0		& 683.17	& 1.03		    & 1.02		    \\
685				& 30.5		& 679.24	& 1.82		    & 1.81		    \\
\hline
\end{tabular}
\label{table:qeff}
\end{table}
\end{center}
\begin{center}
\begin{table}[h]
	\resizebox{\textwidth}{!}{%
\caption{The measured cross sections from the LEDEX experiment in the first diffraction minimum of $^{12}$C along with the Fourier-Bessel (FB) parameterization.}
\begin{tabular}{c|c|c|c|c|c}
\hline
$E_e$		& $\sigma_{exp}$	& $\Delta \sigma_{stat}$	& $\Delta \sigma_{sys}$	& $\sigma_{stat}^{FB}$	& $\sigma_{exp}/\sigma_{stat}^{FB}- 1 $	\\
(MeV)	& (fm$^2$/sr )		& (\%)				& (\%)				& (fm$^2$/sr )		&  (\%)					\\
\hline
362		& $3.26 \times 10^{-8}$			& 7.70		& 3.50		& $3.12 \times 10^{-8} $			& 4.49			\\
685		& $2.35 \times 10^{-7}$			& 4.24		& 2.40		& $1.93 \times 10^{-7}$			& 21.76			\\
\hline
\end{tabular}
\label{table:xsec}
}
\end{table}
\end{center}

\begin{figure}[!htbp]
	\centering
		\includegraphics[width=1.1\textwidth]{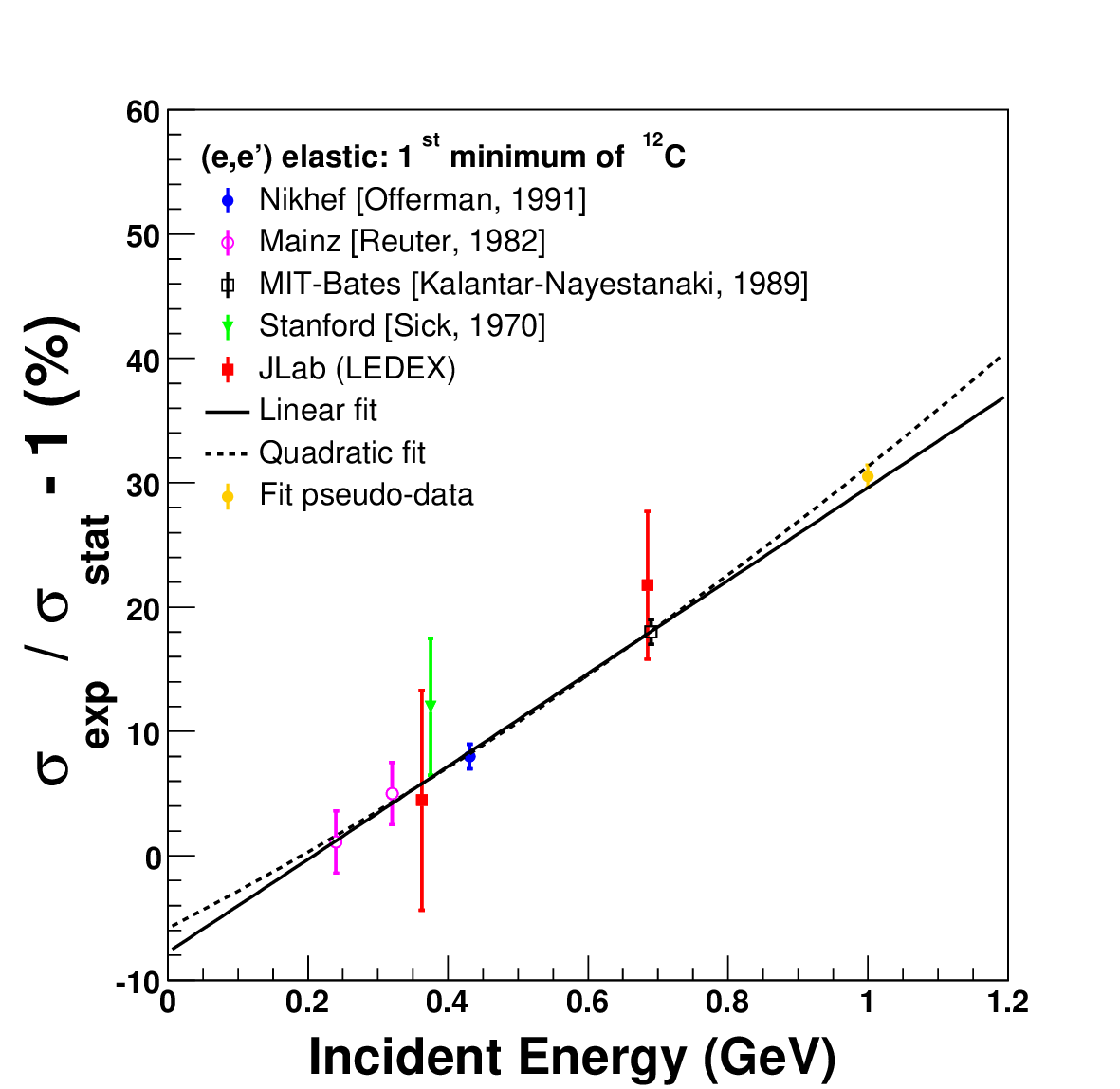}
	\caption{\label{fig:dispersive_results}\protect World data on the energy dependence of dispersive effects in the first diffraction minimum of $^{12}$C. In the y-axis, $\sigma ^{FB}_{stat}$ was replaced by $\sigma _{stat}$ to keep coherency in the text. The first minimum at $q_{eff}=1.84~{\rm fm}^{-1}$ moves slightly with beam energy as 
    noted in~\protect\cite{KalantarNayestanaki:1989nb} (this dependency is out of the scope of this paper). }
\end{figure}

The results of this analysis were also compared to the world data (see Fig.~\ref{fig:dispersive_results}. Note that $\sigma ^{FB}_{stat}$ is replaced by $\sigma _{stat}$ to keep the text coherent throughout this document). From a first order (solid line) and a second order (dashed line) polynomial fits (see Table~\ref{table:fit}), extrapolations {\color{black}indicate} deviations at 1~GeV of 28.9\% and 32.2\%, respectively (average of 30.6\%). One pseudo-data point from the average of the fit functions is also shown at 1~GeV with a 3\% error bar (which is a reasonable systematic error for an elastic peak cross section measurement at Jefferson lab for this energy).

\begin{center}
\begin{table}[hbt]
\caption{Polynomial fit parameters on the world data set for dispersive effects in the first minimum of $^{12}$C.}
\begin{tabular}{l|r|r}
\hline
							& Linear Fit 			& Quadratic Fit  		\\
\hline
$p_0$						& $-6.64  \pm 1.13$	& $-4.40 \pm 4.04 $	\\
$p_1 (10^{-2} ~{\rm MeV}^{-1}$)	& $+3.55 \pm 0.26$	& $+2.36 \pm 2.078$	\\
$p_2 (10^{-5} ~{\rm MeV}^{-2}$)	&				& $+1.30 \pm 2.25$	\\
\hline
$\chi ^2/ndf$					& 2.092/6				&  1.758/5				\\
\hline
\end{tabular}
\label{table:fit}
\end{table}
\end{center}

\begin{figure}[!htbp]
	\centering
		\includegraphics[width=1.\textwidth]{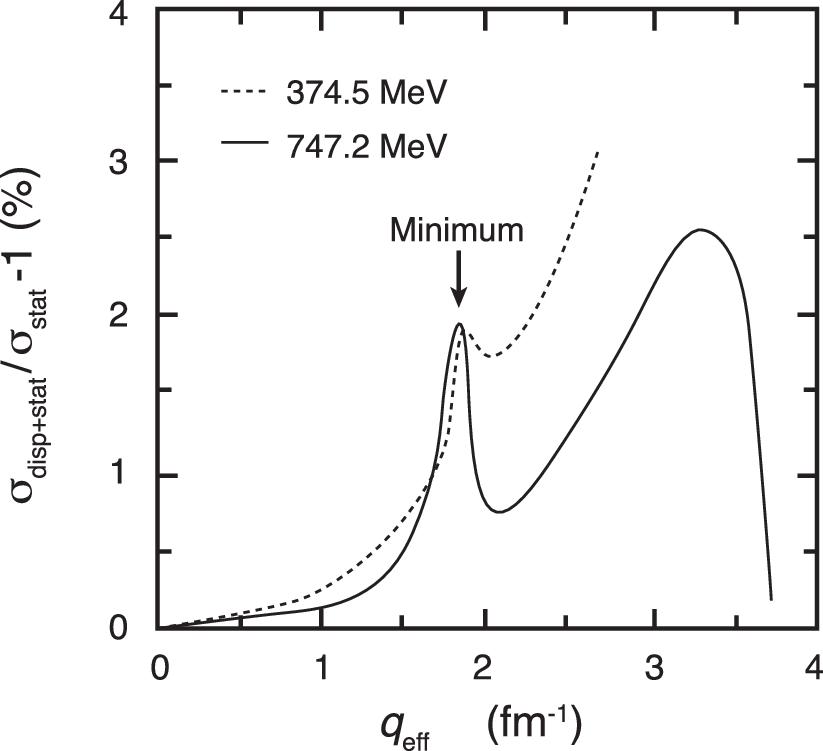}
	\caption{\label{fig:FriarRosen}\protect Calculations of Friar and Rosen~\protect\cite{Friar:1974bn} for dispersion corrections to elastic electron scattering from $^{12}$C at 374.5 and 747.2~MeV in the first diffraction minimum $q_{eff} = 1.84~{\rm fm}^{-1}$.}
\end{figure}

The theoretical prediction from Friar and Rosen~\cite{Friar:1974bn} on the size of dispersive effects in the first diffraction minimum of $^{12}$C is shown in Fig.~\ref{fig:FriarRosen} for 374.5~MeV and 747.2~MeV where the inclusion of dispersive corrections $\sigma _{stat+disp}$ is compared to the cross section $\sigma _{stat}$ obtained from a static charge distribution: the expected (constant) 2\% predicted discrepancy is clearly not reproducing the magnitude and energy dependence behavior seen in the data.

\section{\label{sec:disp_effect}Dispersive corrections and the nuclear matter}

\noindent
A very simplistic approach is now used to estimate the effects of dispersive corrections with our linear and quadratic fits on two specific observables: the nuclear charge density~\cite{Horowitz:2012tj,Abrahamyan:2012gp} and the Coulomb Sum Rule~\cite{Morgenstern:2001jt}.

Coulomb corrections stem from multi-photons exchange between the incoming lepton probe and the target nucleus, with $2\gamma$ being the dominant contribution from higher powers of the $Z\alpha$ terms (with the electromagnetic coupling constant $\alpha = 1/137$). To accurately estimate these effects, one should take into account the continuous change of the incident beam energy while the particle is approaching the nucleus. In practice, one assumes a constant Coulomb field to estimate these effects and applies an effective global shift of the incident and outgoing beam energies as described in Section~\ref{sec:intro}. Note that one should use the averaged Coulomb potential $|V_C|=\int\rho(r)|V_C|(r)d^3r/Z|e|$ instead of the potential at the origin of the nucleus $|V_C(0)|$~\cite{Gueye:1999mm}.

The dispersive cross section $\sigma _{disp} = \sigma _{stat+disp}$ (for simplicity) can be expressed as a function of the cross section $\sigma _{stat}$:
\begin{equation}\label{eq:xsec_disp}
\sigma _{disp} = \sigma _{stat} [ 1 + \delta _{disp} (E_e) ]
\end{equation}
\noindent
with $\delta _{disp} (E_e)$ the higher order correction to the Born Approximation. {\color{black} Our convention throughout the text is to label any quantity with the subscript $disp$, such as the cross section $\sigma_{disp}$, that has been directly obtained from experimental measurements and is affected by the contribution from dispersive effects. Analogously, the subscript $stat$, such as $\sigma_{stat}$, is attached to any quantity that could be obtained by removing the contribution from dispersive effects, thus correcting the experimental observation. In that sense $\sigma_{stat}$ will be the expected cross section from the Born Approximation. Equation~\eqref{eq:xsec_disp} states that the observed experimental cross sections $\sigma_{disp}$ could be modeled by a small multiplicative perturbation added to the static $\sigma_{stat}$ cross section.}

	\subsection{\label{sec:nuclear_size}Effects on nuclear radii}

\noindent
In the Plane Wave Born Approximation, the nuclear charge density distribution $\rho_{\rm ch}(r)$ is the Fourier transform of the nuclear form factor and for spherically symmetric charge distributions the relation is~\cite{deVries87}:
\begin{equation}\label{eq:rho_FF}
\rho_{\rm ch}(r) = \frac{1}{2\pi^2}\int F_{\rm ch}(q)\frac{\sin(qr)}{qr}q^2dq
\end{equation}

$\rho_{\rm ch}(r)$ can thus be extracted from the experimentally measured $F_{\rm ch}(q^2)$ and it is usually normalized to either $1$ or the total charge of the nucleus. We adopt the first convention in this work:
\begin{equation}\label{eq:rho_norm}
4\pi \int \rho_{\rm ch}(r)r^2dr = 1
\end{equation}

A model independent analysis can be done to extract the nuclear charge density distributions using either a sum of Gaussian (SOG)~\cite{Sick1974509} or sum of Bessel (FB)~\cite{Dreher1974219} functions. We will only focus on the latter and refer the readers to reference~\cite{deVries87} for more details on the former.

One can use the zero'th spherical Bessel function $j_0(r) = \sin(qr)/qr$ to expand the charge density as:
\begin{equation}\label{eq:rho_FB}
\rho_{\rm ch}^{\rm FB}(r) =
\left\{
\begin{array}{cll}
\sum _\nu a_\nu j_0\big (\frac{\nu \pi r}{R_{\rm cut}}\big )	& {\rm for} & r \leq R_{\rm cut} \\
									&  		& \\
0									& {\rm for} & r> R_{\rm cut}
\end{array}
\right.
\end{equation}
with $R_{\rm cut}$ the cut-off radius chosen such as the charge distribution is zero beyond that value ($R_{\rm cut} = 8$~fm for $^{12}$C~{\color{black}\cite{Offermann:1991ft}}) and the coefficients $a_\nu$ related to the form factor as $a_\nu = q_\nu^2 F_{\rm ch}(q_\nu)/2 \pi R_{cut}$, where $q_\nu = \nu \pi /R_{cut}$ is obtained from the $\nu$-th zero of the Bessel function $j_0$.

In this study we will ignore the contribution of the neutrons to the electric charge distribution of the nucleus\footnote{Even though the neutron has a total electric charge of zero, its charge density $\rho_n(r)$ is not zero. {\color{black}Nevertheless,} its contribution to the total charge density of the nucleus is small.}. Therefore, $\rho_\text{ch}(r)$ could be considered as resulting from folding the distribution $\rho _{\rm nuc}(r)$ of the nucleons, protons in our approximation, inside the nucleus with the finite extension of the protons $\rho _p(r)$~\cite{Dreher1974219}. The Fourier transform of $\rho _{\rm ch}(r)$ is then given by the product of the transform of $\rho _{\rm nuc}(r)$ and $\rho _p(r)$:
\begin{equation}\label{eq:FF_ch_nuc_p}
F_{\rm ch}(q) = F_{\rm nuc}(q)F_p(q)
\end{equation}

The relationship between the corresponding radii is:
\begin{equation}\label{eq:radii}
R_{\rm ch}^2 = R_{\rm nuc}^2 + R_p^2
\end{equation}
\noindent
with {\color{black}$R_p = 0.8414(19)~{\rm fm}$} the proton radius{\color{black}~\cite{CODATA18}}. The rms $\langle r^2_{\rm ch}\rangle^{1/2}$ can then be obtained from the nuclear charge density distribution ($\rho_{\rm ch}$) which extends up to $R_{\rm cut}$. Its general expression is:

\begin{equation}\label{eq:rms}
\langle r_{\rm ch}^2 \rangle = \int _0 ^{R_{\rm cut}} \rho_{\rm ch}(r) r^2 d^3r = 4\pi \int _0 ^{R_{\rm cut}} \rho_{\rm ch}(r) r^4 dr
\end{equation}

Using the Bessel expansion of $\rho_{\rm ch}$ from Eq.~\eqref{eq:rho_FB} leads to:
\begin{equation}\label{eq:rms_bessel}
\langle r_{\rm ch}^2 \rangle = 4\pi \int_0^{R_{\rm cut}} \sum _\nu a_\nu j_0 \left( \frac{\nu\pi r}{R_{\rm cut}} \right) r^4 dr
\end{equation}

Evaluating the integral of the Bessel function gives:
\begin{equation}\label{eq:integral_bessel}
\int_0^{R_{cut}} j_0 \left( \frac{\nu\pi r}{R_{cut}} \right) r^4 dr = \frac{(-1)^{\nu}R_{cut}^5(6-\nu^2\pi^2)}{\nu^4\pi^4} 
\end{equation}

Substituting into Eq.~\eqref{eq:rms}:
\begin{equation}\label{eq:rms_Rcut}
\langle r_{\rm ch}^2 \rangle = 4\pi \sum_\nu a_\nu  \frac{(-1)^{\nu}R_{\rm cut}^5(6-\nu^2\pi^2)}{\nu^4\pi^4}
\end{equation}

Therefore, all the coefficients $a_\nu$ of the Fourier Bessel expansion play a role in estimating the radius of the charge density distribution, decreasing in importance as $1/\nu^2$. If the measured cross sections used to extract the value of the form factor $F_{\rm ch}(q)$ are indeed modified by the dispersive corrections, then the change would propagate through the fitted coefficients $a_\nu$ to the estimate of the charge radius $R_{\rm ch}\equiv \langle r_\text{ch} ^2 \rangle ^{1/2}$. The total change in $R_{\rm ch}$ can be written as (see Appendix ~\ref{appendix:betacalc} for details): 

\begin{equation}\label{eq: Ap del Rch (MainTxt)}
 \delta R_{ch} = \sum _i ^N \frac{\partial R_{ch}}{\partial y_i} \delta y_i = \sum _i ^N \Big(\sum_\nu ^M \frac{\partial R_{ch}}{\partial a_\nu} \frac{\partial a_\nu}{\partial y_i} \Big) \delta y_i,
\end{equation}

\noindent
where $\delta y_i$ is the change in the ${\rm i^{th}}$ value of the form factor $y_i = F(q_i)$, in this case due to the dispersive effects. Estimating the exact values of $\delta y_i$ is a complicated task beyond our scope since the change in the cross section as shown in Eq.~\eqref{eq:xsec_disp} depends on the energy, but the momentum transfer $q$ is a function of both the energy and the angle $\theta$. Therefore, for the same fixed value of $q$ we could have different pairs of $(E,\theta)$ which will be impacted differently. 

In order to simplify our discussion, we assume that we can separate the total effect of the dispersive effects on the form factor values as:
\begin{equation}\label{eq: dispF}
    F_{disp}(q) =F(q)_{stat} [ 1 + \frac{1}{2}\delta (E_e)S(q) ],
\end{equation}

\noindent 
with {\color{black}$\delta _{disp} = \delta (E_e)S(q)$} from Eq.~\eqref{eq:xsec_disp} where $\delta(E_e)$ controls the overall strength of the perturbation and $S(q)$ controls the impact this change would have on different $q$ values. The factor of $1/2$ comes from assuming that $\delta(E_e)$ is small and propagating the change from Eqs.~\eqref{eq:xsec_theo} and~\eqref{eq:xsec_disp}: $F\propto \sqrt{\sigma}$ which implies $\delta F / F \propto (1/2) \  \delta \sigma / \sigma$.

{\color{black} Since the variable $q$ depends on both $E_e$ and $\theta$, a separation such as Eq.~\eqref{eq: dispF} might not be completely accurate. As it can be seen in the calculations of Friar and Rosen (Fig.~\ref{fig:FriarRosen}), a change in $E_e$ clearly affects the overal shape of the dispersion corrections as a function of $q$. Nevertheless, Eq.~\eqref{eq: dispF} is simple enough to allow providing an estimate for the impact of such a change in inferred nuclear properties of the nucleus. In particular, we can write the change in the charge radius as:}

\begin{equation}
  R_{\rm ch}^{disp} = R_{\rm ch} ^{stat} \left[ 1 + \beta \delta(E_e)     \right] .
\end{equation}

\noindent
where $\beta$ is a proportionality coefficient fixed once $S(q)$ is specified {\color{black}(for a given fixed strength $\delta(E_e)$, the change in the radius will depend on the shape of $S(q)$, which is encoded in $\beta$)}. Table~\ref{ResultsTable} shows the results {\color{black}(see the Appendix for a detailed description)} for three different test perturbations $S(q)$  {\color{black} plus an empirical one}, when using the data  {\color{black} without dispersive corrections }from Offermann~\cite{Offermann:1991ft} ({\color{black} Table X}) for the central values of the form factor.  {\color{black} For the three test cases} these values were modified assuming a constant high value of {\color{black} $\delta$}$ (E_e) = 30\%$. 

The forms for $S(q)$ were divided into two categories: $\delta_4$ and $\delta_5$ represent {\color{black}up}{\color{black}-shift} of {\color{black} 1 ($15\%$ when multiplied by $1/2 \ \delta(E_e)$)} on the value of $F(q_\nu)$ for $\nu=4$ and $\nu=5$, respectively, while Gaussian represents a Gaussian {\color{black} up}-shift of amplitude {\color{black} 1 at its peak (once again $15\%$ when multiplied by $1/2 \ \delta(E_e)$)}, centered at the diffraction minimum $q=1.84$ fm$^{-1}$ and with a standard deviation of $0.25$ fm$^{-1}$. {\color{black} An overall up-shift in the form factor was chosen based on the calculations shown on Fig.~\ref{fig:FriarRosen}, which predict an up-shift in the observed cross sections due to the dispersive effects, which means $\sigma_{disp} \geq \sigma_{stat} $.}

{\color{black} The empirical perturbation was obtained as  $\delta_{emp}(q_\nu)=[F^*_{disp}(q_\nu)-F^*_{stat}(q_\nu)] / F^*_{stat}(q_\nu)$ , where $F^*_{disp}(q_\nu)$ ($F^*_{stat}(q_\nu)$) represents the form factor values obtained from the second (third) column in Table X of \cite{Offermann:1991ft}. Since no amplitude $\delta (E_e)$ was involved in the empirical perturbation, the value of $\beta$ cannot be defined and we have that:}

\begin{equation}
    {\color{black}F_{disp}(q_\nu) =F(q_\nu)_{stat} [ 1 + \delta_{emp}(q_\nu) ].}
\end{equation}

\begin{table}[hbt]
\caption{The first column shows the perturbation form $S(q)$ {\color{black}in addition to the empirical perturbation}. In {\color{black} the first three} cases a strength of {\color{black}$\delta$}$ (E_e)=0.3$ was assumed. The second column shows the calculated new radius, {\color{black}$R_\text{ch}^{stat}$} (the original radius is $2.4711$ fm). The third and fourth columns show the $\beta$ coefficient and the percentage change in $R_\text{ch}$, {\color{black}namely $\delta R_\text{ch} \equiv R_\text{ch}^{stat} -R_\text{ch}^{disp}$,} respectively.}
\label{ResultsTable}
\begin{tabular}{c|c|c|c}
\hline
\multirow{2}{*}{$S(q)$} & R$_\text{ch}^{\color{black}{{stat}}}$   & \multirow{2}{*}{\color{black}$\beta$} & \multirow{2}{*}{$\delta$ R$_\text{ch}$} \\
                        &  [\text{fm}]                          &                                       &                   \\
\hline
$\delta_4$    & 2.512                     & -0.055         & {\color{black} 1.65} \%           \\
$\delta_5$    & 2.480                     & -0.012         & {\color{black} 0.35} \%           \\
Gaussian    & 2.495                     & -0.032         & {\color{black} 0.98} \%           \\
{\color{black} Empirical}    & {\color{black} 2.477  }                   & -      & {\color{black} 0.25} \%           \\
\hline
\end{tabular}
\end{table}

Therefore, while the fits parameters from Table~\ref{table:fit} imply corrections expected to be around 30\% on the cross section at 1~GeV for $^{12}$C, the effect on the nuclear charge radius from our test calculations is around a percent. A detailed analysis of the impact of dispersive effects on nuclear radii was performed by Offermann et al.~\cite{Offermann:1991ft}: the result is a net relatively small effect of 0.28\%, implying a renormalization of the charge distribution to offset the change in the cross section. 

{\color{black} When using the empirical perturbation for the $\delta y_i$ in Eq.~\eqref{eq: Ap del Rch (MainTxt)}  we obtain an effect of $0.25\%$ in the radius, very close to the actual $0.26\%$ (reported as $0.28\%$ when using rounded values for the radii) in \cite{Offermann:1991ft}. It seems that the strength ($30\%$) of the other three perturbations is too big to reproduce the small change in the radius, which might indicate that the effects \emph{on the available data} of the dispersive corrections are roughly at least a factor of five smaller outside the vicinity of the difraction minimum.}

The Coulomb field extracted from $\langle r^2 \rangle ^{1/2}$ should then also be modified from
\begin{equation}\label{eq:field_disp}
\mid V_C \mid~=~\mid V ^{stat}_C \mid~=~\frac{KZ}{\langle r^2 \rangle ^{1/2} } \; ; K = 1/4\pi\varepsilon_0
\end{equation}
to
\begin{equation}\label{eq:VC_disp}
\mid V ^{disp}_C \mid~=~\mid V ^{stat}_C \mid / [ 1 + \beta \delta (E_e) ] \;
\end{equation}

As mentioned previously, Coulomb corrections are expected to be comparatively small for GeV energies: $S_{HOB} =  2.6\%$ for a 1~GeV incident electron beam on a $^{208}$Pb target. In the remainder of this section, we will assume that the energy dependent correction is solely rising from dispersive corrections and is embedded in the term $\delta _{disp} (E_e)$.

In order to estimate the corrections for $^{208}$Pb, we scale the carbon value using Coulomb fields from~\cite{Gueye:1999mm}:

\begin{itemize}
\item The scaling is first calculated from the super ratio:
\begin{equation}\label{eq:scale_disp_effect}
R_{\rm scale} = \frac{V_{C, {\rm ^{208}Pb}} =18.5~{\rm MeV}}{V_{C, ^{12}{\rm C}} = 5.0~{\rm MeV}} \frac{Z_{\rm ^{12}C} = 6}{Z_{\rm ^{208}Pb} = 82}  = 26.34\%
\end{equation}
Thus giving a value for the dispersive corrections of $26.34\% \times 30\% \simeq 8\%$ that is compatible with the $\sim 6\%$ effect observed by Breton et al.~\cite{Breton:1991fe}. 
\item The effect on the lead radius can then be obtained by applying the above scaling to the value from Offermann et al.~\cite{Offermann:1991ft}
\begin{equation}\label{eq:disp_effect}
0.28\% R_{\rm scale} = 0.07\%.
\end{equation}
The reported experimental value of the charge radius of lead is~\cite{Angeli201369} $R_{\rm ch} = 5.5012(13)$~fm which would imply an upward shift to 5.5053(13) fm when taking the 0.07\% scaling into account. 
\end{itemize}

The situation is far more complex for parity-violating experiments~\cite{Horowitz:2012tj,Abrahamyan:2012gp,Abrahamyan:2012cg} from which the measured asymmetry is used to extract a neutron skin. These experiments typically occurred near diffractive minima to maximize their sensitivity to the physics~\cite{Piekarewicz16}, where also dispersive corrections contribute the most. Our estimation suggests the importance of this correction for high precision determinations of the radius and/or the neutron skin of heavy nuclei.

It is clear one should take dispersive effects into account; however, to our knowledge, there is no known measurements of dispersive effects using polarized beams and/or target. Therefore, measurements of the energy dependence for dispersive effects using polarized elastic scattering on various nuclear targets ($A>1$) should be performed to provide an accurate information about the size of these effects in and outside minima of diffraction. 

	\subsection{\label{sec:csr}Possible effects on the Coulomb Sum Rule}

\noindent
The Coulomb Sum Rule (CSR)~\cite{McVoy:1962zz} is defined as the integral of the longitudinal response function $R_L(\omega, |{\bf q}|)$ extracted from quasi-elastic electron scattering:
\begin{equation}\label{eq:CSR}
S_L(|{\bf q}|) = \int_{\omega>0}^{|{\bf q}| }\frac{R_L(\omega, |{\bf q}|)}{ZG_{E_p}^2(Q^2) + NG_{E_n}^2(Q^2)}d\omega
\end{equation}
where $-Q^2 = \omega^2 - \vec{q}^2$ with $\omega$ the energy transfer and $\vec{q}$ the three-momentum transfer. $G_{E_{p,n}}(Q^2)$ is the  proton (neutron) form factor which reduces to the Sachs electric form factor if the nucleon is not modified by the nuclear medium~\cite{Noble:1980my}. $\omega >0$ ensures that the integration is performed above the elastic peak. In essence, CSR states that by integrating the longitudinal strength over the full range of energy loss $\omega$ at large enough momentum transfer $q$, one should get the total charge (number of protons) of a nucleus.

The quenching of CSR has been found to be as much as 30\%~\cite{Morgenstern:2001jt} for medium and heavy nuclei. Using a quantum field-theoretic quark-level approach which preserves the symmetries of quantum chromodynamics, as well as exhibiting dynamical chiral symmetry breaking and quark confinement, the most recent calculation by Cloet et al.~\cite{Cloet:2015tha} confirmed the dramatic quenching of the Coulomb Sum Rule for momentum transfers ${\mid} q {\mid}{\gtrsim}2.5~{\rm fm^{-1}}$ that lies in changes to the proton Dirac form factor induced by the nuclear medium. 

\begin{figure}[!htbp]
	\centering
		\includegraphics[width=1.\textwidth]{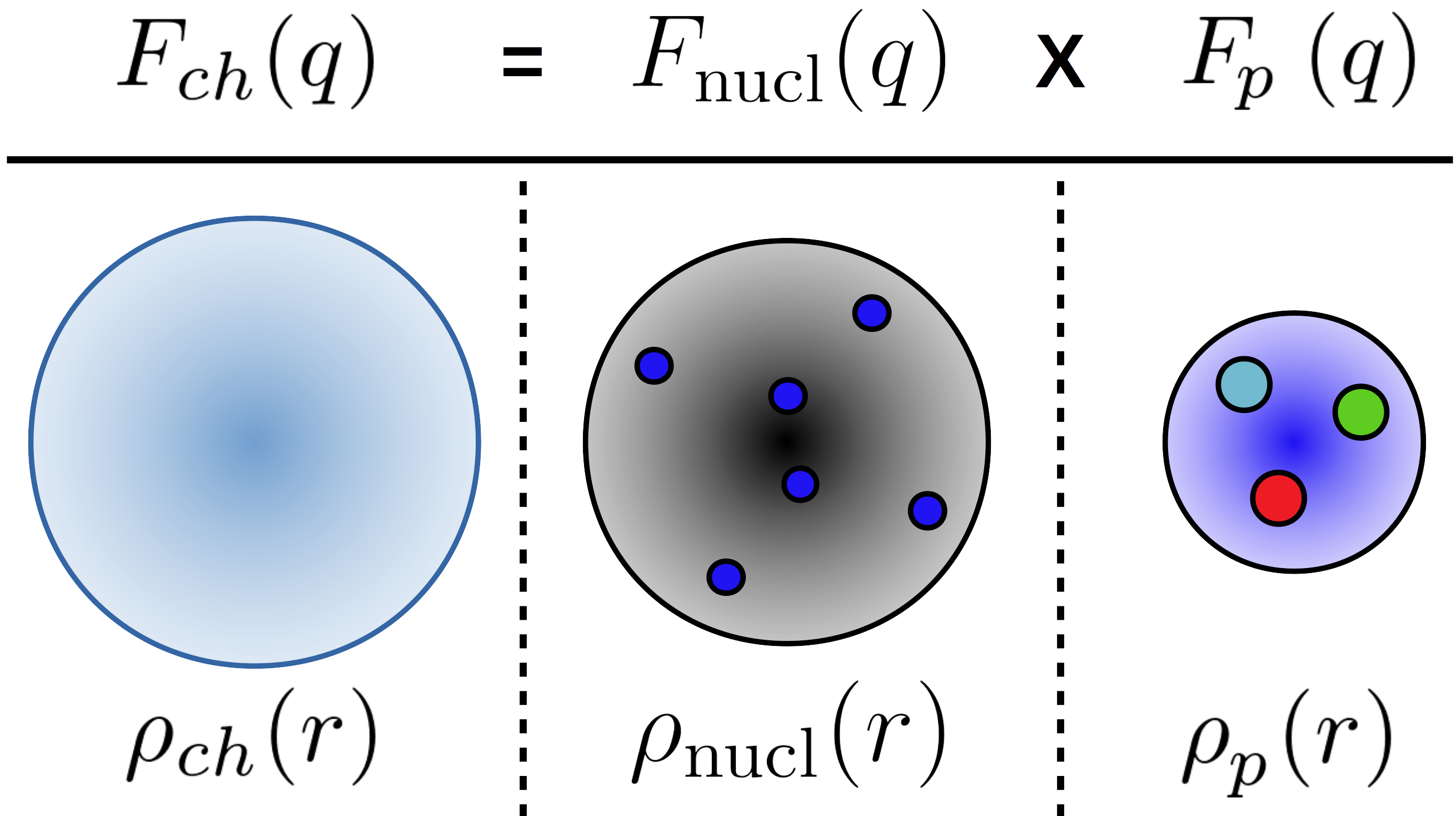}
	\caption{\label{fig:ChargeDistributions}\protect Relationship between the charge, nucleons (protons) and the single proton form factors along with their respective densities for $^{12}$C. The protons density $\rho_\text{nucl}$ specifies the spatial distribution of the 6 protons inside the $^{12}$C nucleus, treating them as point particles (blue circles over the black background in the middle column). The charge form factor $F_{ch}$, which relates to the \emph{charge} distribution in the nucleus (left column), is the result of folding the protons form factor $F_\text{nucl}$ with the single proton form factor $F_p$, which relates to the charge distribution inside the proton (right column, the color circles represents the three quarks).  }
\end{figure}

As previously noted, the nuclear charge distribution $\rho_{\rm ch}(r)$ may be considered as a result from folding the distribution $\rho_{\rm nuc}(r)$ of the nucleons in the nucleus with the finite extension of the nucleons $\rho_p(r)$~\cite{Dreher1974219} as represented in Fig.~\ref{fig:ChargeDistributions}.


Quasi-elastic electron scattering corresponds to a process in which electrons elastically scattered off nucleons. The nuclear response is affected by the fact that nucleons are not free and carry a momentum distribution, the existence of nucleon-nucleon interactions and interactions between the incoming and outgoing probe and recoils. Therefore, noting that $R_L$ probes $\rho_{\rm nuc} = \rho_{\rm protons}$ while elastic scattering experiments probe $\rho_{\rm ch}(r)$, any measured shift of $F_{ch}(q)$ results from a change in $F_{\rm nuc}$ or $F_p$, or both. Even when considering {\color{black}the contribution from two-photon exchanges that are responsible for the measured deviation between unpolarized and polarized electron scattering in the extraction of the $\mu G^p_E/G^p_M$ ratio and also believed to be at the origin of the proton {\color{black}form factor} puzzle~\cite{Blunden:2005ew}}  (see the Introduction section), the discrepancy observed cannot explain the 30\% quenching of $R_L$~\cite{Carlson:2007sp,Arrington:2011dn,Afanasev:2017gsk}. In the following, we assume that the contribution from dispersive effects found in $\rho_{\rm ch}(r)$ translates entirely in a change in $\rho_{\rm protons}$ and hence in the CSR.

From our naive model ({\color{black}with nuc = p or n}):
\begin{equation}\label{eq:GE}
G^{disp}_{E_{\rm nuc}}(Q^2) = \frac{G^{stat}_{E_{\rm nuc}}(Q^2)}{1 + \beta \delta(E_e)}
\end{equation}
Hence:
\begin{equation}\label{eq:CSR_disp}
S_L^{disp}(|{\bf q}|) = S^{stat}_L(|{\bf q}|)~\times~[ 1 + \beta \delta(E_e)]
\end{equation}

Using Fig.~\ref{fig:dispersive_results} for a 600~MeV incident beam on $^{12}$C, one would expect a 15\% correction in the minimum of diffraction, which is a factor of 7.5 from the 2\% prediction from Friar and Rosen~\cite{Friar:1974bn}. Above the minimum, their prediction indicates an almost linear increase of the dispersion corrections up to about 3.3~fm$^{-1}$ where it reaches a maximum of about 3\%. Assuming the same scaling, that is a $0.03 \times 7.5 \simeq 22\%$ predicted effect in the kinematic regime of the CSR data for $^{12}$C~\cite{Barreau:1983ht}. Therefore, dispersion corrections could have a significant contribution on the CSR quenching if the experimentally measured longitudinal response function $R_L(\omega, |{\bf q}|)$ is corrected for these effects. 

\section{\label{sec:conclusion}Conclusion\protect}

We have presented new results on the energy dependence for \textit{dynamic} dispersion corrections in elastic electron scattering in the first diffraction minimum of $^{12}$C 
at $q \approx 1.84~{\rm fm}^{-1}$ from Jefferson Lab obtained at two different energies: 362~MeV and 685~MeV~\cite{Ledex}. The results are in very good agreement with previous world data on this topic and cannot be explained with available theoretical calculations.

{\color{black}We presented a general theoretical framework that allows to propagate the dispersive correction effects, treated as a perturbation, to the coefficients of a Bessel function fit of the form factor. We first benchmarked our calculation using the experimental data on $^{12}$C from Offermann et al.~\cite{Offermann:1991ft}:} we investigated the impact of these corrections on the nuclear charge density radius and obtained comparable results with the ones reported by the authors. Using scaling arguments, we {\color{black} then} find this contribution to be around 0.07\% for the recent measurement of the nucleon radii from Pb~\cite{Horowitz:2012tj,Abrahamyan:2012gp,Abrahamyan:2012cg}. While we find this contribution to be relatively small, {\color{black}it will take a detailed investigation and theory to understand how this affects the parity-violating asymmetry. A subsequent study on the observed quenching of the Coulomb Sum Rule~\cite{Cloet:2015tha} indicates that the expected contribution seems to be larger.}

{\color{black}Note that from the analysis presented here, nothing precludes dispersive effects for being zero or even having a different sign on some measured observables.} Therefore, we conclude it is important that a systematic study of the dispersion corrections inside and outside diffraction minima for a large range of (light through heavy) nuclei be performed using both unpolarized and polarized beams/targets to help provide a more complete understanding of elastic (and inelastic) electron/positron-nucleus scattering.

\section*{Acknowledgements}
\noindent
We thank Larry Cardman for many useful discussions.
This work was supported by the U.S. Department of Energy National Nuclear Security Administration under award number {DE-NA0000979},  by the U.S. Department of Energy grant DE-AC02-06CH11357, by the U.S. National Science Foundation grant NSF-PHY-1505615 and by the U.S.  Department of Energy contract DE-AC05-06OR23177 under which Jefferson Science Associates operates the Thomas Jefferson National Accelerator Facility.

\begin{appendix}
\section{Propagation of changes from the Form Factor to the charge radius}\label{appendix:betacalc}

\subsection{Formalism}

We are interested in estimating how a change in the observed cross section, or the deduced form factor values, could impact the extracted radius $R_{ch}$. 

The charge radius is a function of the $M$ parameters of our model~\eqref{eq:rms}, in this case the $M$ independent Bessel coefficients $a_\nu$, which in turn depend on the $N$ experimentally extracted form factor values $y_i$. Therefore, through the coefficients $a_\nu$ the charge radius is a function of the experimental points and one can write a small change in $R_{ch}$ due to a given small change in the observations $(\delta y_1, \delta y_2, ... , \delta y_N)$ as:

\begin{equation}\label{eq: Ap del Rch}
 \delta R_{ch} = \sum _i ^N \frac{\partial R_{ch}}{\partial y_i} \delta y_i = \sum _i ^N \Big(\sum_\nu ^M \frac{\partial R_{ch}}{\partial {\color{black}a_\nu}} \frac{\partial a_\nu}{\partial y_i} \Big) \delta y_i
\end{equation}

For $M$ independent coefficients $a_\nu$, one has $M+1$ Bessel functions in our model due to the normalization constraint. The $a_{M+1}$ can be explicitly written by solving the constraint:

\begin{equation}\label{eq: Ap Normalization}
    \left.
    \begin{array}{l}
    {\color{black}4\pi \int \rho(r) r^2 dr = 1} , \\
    \\
    \sum_\nu^{M+1} (-1)^{\nu+1}\frac{4\pi R_{cut}}{q_\nu^2}a_\nu  =1, \\
    \\
    a_{M+1} = (-1)^{M} \Big(1- \sum_\nu^{M} (-1)^{\nu+1}\frac{4 \pi R_{cut}}{q_\nu^2}a_\nu\Big) \frac{(M+1)^2 \pi}{4 R_{cut}^3}
    \end{array}
    \right.
\end{equation}

An alternative route would be to use Lagrange multipliers when making calculations for the data fit, which would allow to treat the $M+1$ coefficients independently. Following Eq.~\eqref{eq:rms_Rcut}, and taking into account the normalization condition, the partial derivative of $R_\text{ch}$ with respect to a coefficient $a_\nu$ is given by:

\begin{equation}
 \frac{\partial R_{ch}}{\partial a_\nu} =  \frac{1}{2R_{ch}} 4\pi\frac{(-1)^\nu R^5 _{cut} (6-\nu^2\pi^2)}{\nu^4\pi^4} + \frac{\partial R_{ch}}{\partial a_{M+1}}\frac{\partial a_{M+1}}{\partial a_\nu}
\end{equation}

The last term has to be included since $R_\text{ch}$ depends on the $M+1$ coefficients and $a_{M+1}$ depends linearly on the rest of the $a_\nu$, making the calculation straightforward from Eq.~\eqref{eq: Ap Normalization}.

Meanwhile, the change in the coefficient $a_\nu$ due to a change in $y_i$ is a little more challenging to compute. To do so, one must specify how exactly the coefficients where obtained from the experimental data. An usual way is by minimizing the sum of the squares of the residuals denoted by $\chi^2$:

\begin{equation}
 \chi^2 \equiv \sum_i ^N \frac{[F(q_i,\vec{a})-y_i]^2}{2 \Delta y_i^2},
\end{equation}

\noindent
where $\Delta y_i$ is the estimated error, or uncertainty, in the measurement $y_i$ and $\boldsymbol{a}$ is the list of coefficients $a_\nu$. The optimal values of the parameters $\vec{a}_{opt}$ is found by imposing the condition of a minimum:

\begin{equation}
 \frac{\partial \chi^2}{\partial a_\nu} \Big|_{a_{opt}} \equiv G_\nu(\vec{a},\vec{y}) \big|_{a_{opt}} =0
\end{equation}

Now, the key point is that one has $M$ different $G_\nu$ which are functions of the parameters $\vec{a}$ and the observations $y_i$, and they all equal zero when evaluated at the optimal parameters $\vec{a}_{opt}$. If the value of one observation $y_i$ changes by a small amount $\delta y_i$, the minimum of $\chi^2$ will move in the parameter space by a small amount. One can calculate this displacement by noticing that all the parameter values $a_\nu$ would have to change accordingly in order to keep the values of each $G_\nu$ at zero. Quantitatively this implies: $ \frac{\partial G_\nu}{\partial y_i} \delta y_i = - \sum_k ^M \frac{\partial G_\nu}{\partial a_k} \delta a_k$ for $\nu \in(1,... M)$, which can be put in a matrix equation:

\begin{align*}
  \frac{\partial G_1}{\partial y_i}& \delta y_i =  - \Big( \frac{\partial G_1}{\partial a_1}\delta a_1+ \frac{\partial G_1}{\partial a_2} \delta a_2 \  ... \ + \frac{\partial G_1}{\partial a_M} \delta a_M  \Big) \\
   \frac{\partial G_2}{\partial y_i}& \delta y_i = - \Big( \frac{\partial G_2}{\partial a_1}\delta a_1+ \frac{\partial G_2}{\partial a_2} \delta a_2 \  ... \ + \frac{\partial G_2}{\partial a_M} \delta a_M  \Big)  \\
   &\;\;\vdots \notag \ \ \ \ \ \ \ \ \ \ \ \  \ \ \ \ \ \ \ \ \ \ \ \  \;\;\vdots \notag \\
    \frac{\partial G_M}{\partial y_i}& \delta y_i = - \Big( \frac{\partial G_M}{\partial a_1}\delta a_1+ \frac{\partial G_M}{\partial a_2} \delta a_2 \  ... \ + \frac{\partial G_M}{\partial a_M} \delta a_M  \Big) 
\end{align*}

\noindent
resulting in:

\begin{equation}
  \begin{aligned}
  \frac{\partial \vec G}{\partial y_i} \delta y_i = - \mathcal{H} \vec{\delta a} \Rightarrow
  \vec{\delta a} = - \Big(\mathcal{H}^{-1} \Big) \frac{\partial \vec{G}}{\partial y_i} \delta y_i,
  \end{aligned}
\end{equation}

Since $G$ was already first derivatives of $\chi^2$ with respect to the parameters, the expression obtained is $\mathcal{H}_{[j,k]}\equiv \frac{\partial ^2 \chi^2}{\partial a_j \partial a_k}$, the Hessian matrix which contains second derivatives of $\chi^2$. From this equation one can finally extract how each parameter $a_\nu$ changes when an observation $y_i$ changes:

\begin{equation}\label{eq: MatDanuDyi}
\frac{\partial a_\nu}{\partial y_i} = - \Bigg[ \Big(\mathcal{H}^{-1} \Big) \frac{\partial \vec{G}}{\partial y_i} \Bigg]_{[\nu]} = -\sum_k^M \mathcal{H}^{-1}_{[\nu,k]} \frac{\partial G _k}{\partial y_i}.
\end{equation}

From the set of changes in the observations, $\delta y_i$, due to the dispersive corrections, one has all the ingredients needed to calculate the change in $R_{ch}$ from Eq.~\eqref{eq: Ap del Rch}. In the following {\color{black}discussion,} we apply this framework to the data set presented by Offermann et al.~\cite{Offermann:1991ft} {\color{black} under the convention that $\delta R_\text{ch} = R_\text{ch}^{stat} -R_\text{ch}^{disp}$, since we want to estimate the change in the radius once the corrections for the dispersive effects have been implemented}. 

\subsection{Example: Change in the nuclear radius of $^{12}\text{C}$}

We use the work from~\cite{Offermann:1991ft} where the authors used 18 Bessel functions to fit cross section experimental data from $^{12}\text{C}$. To show our method, we use the values of their first 9 coefficients $a_\nu$ $\nu \in \{1,9\}$ from their Table X second column (without dispersion corrections) to generate 9 values $y_\nu$ of the form factor according to the relation $a_\nu = F(q_\nu)q_\nu^2/2\pi R_\text{cut}$ at those 9 special $q_\nu$ values with $R_\text{cut}= 8$ fm. For the error associated with each "observation" $y_\nu$, we use the adapted error $\Delta y_\nu$ from their reported percentage error in $\Delta a_\nu$. For the remaining 9 points $\nu \in \{10,18\}$, we center the observations $y_\nu$ at zero and add an error band associated with the form factor of the proton as the authors did following the recommendation in \cite{Dreher1974219}. Since the normalization condition must be respected, only 17 from the 18 coefficients $a_\nu$ are independent. We identify therefore $N=18$ and $M=17$.

\begin{figure}[htbp]
	\centering
		\includegraphics[width=0.8\textwidth]{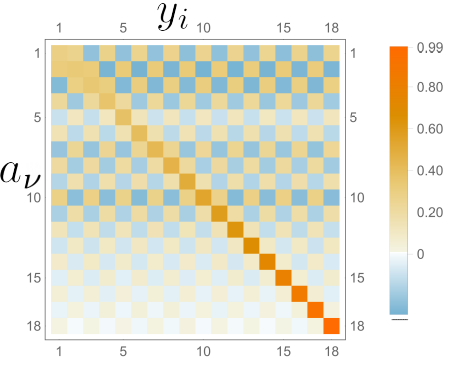}
	\caption{\label{fig:Matrix}  $\partial a_\nu/ \partial y_i$ matrix for the data extracted from Offermann et al.~\cite{Offermann:1991ft}.}
\end{figure}

Figure~\ref{fig:Matrix} shows the matrix $\partial a_\nu/ \partial y_i$ from Eq.~\eqref{eq: MatDanuDyi} for the 18 observations $y_i$ and 17 + 1 coefficients $a_\nu$. Even though we are not treating $a_{18}$ as an independent variable since we solved the constraint explicitly, we can still calculate how much its value changes when any one of the observations $y_i$ changes. It can be seen that as $\nu$ increases, $a_\nu$ becomes more dependent on $y_\nu$ and less sensitive to other values of $y$. In principle, if the 18 coefficients were independent, each $a_\nu$ will only be sensitive to their corresponding $y_\nu$, but the normalization constraint introduces mixing.

In the third column of Table~\ref{TFunctions} are the numerical values of $\partial R_\text{ch}/ \partial y_i$ for the first $9$ observations $y_i$. Each one of these numbers, when multiplied by a small change in their associated observation, will yield the corresponding small change in $R_\text{ch}$ as in Eq.~\eqref{eq: Ap del Rch}. The fourth column shows the percentage change needed in observation $y_i$ to create a 1\% change in the radius. Even though the values $\partial R_\text{ch}/ \partial y_i$ are roughly the same size for all the observations, this fourth column shows that $R_\text{ch}$ is more sensitive to \emph{percentage} changes in the first observations. 

\begin{table}[]
\caption{The first column shows the index number of the special momentum transfer $i\pi/R_\text{cut}$ and the second column its form factor value obtained from \cite{Offermann:1991ft}. The third column shows the value of $\partial R_\text{ch}/ \partial y_i$ . The fourth column shows the percentage change needed in $y_i$ to generate an equivalent change of $1\%$ in the estimated charge radius.}
\label{TFunctions}
\begin{tabular}{c|c|c|c}
\hline
\multirow{2}{*}{Location $i$} & \multirow{2}{*}{$y_i$}  & $\partial R_\text{ch}/ \partial y_i$  \ \ \  & $\delta R_{ch}$ \\
    & & [fm] & $ = 1\%$\\
\hline
1 & 0.854 & -9.214 & 0.3   \\
2 & 0.526 & -2.595 & 1.8    \\
3 & 0.221 & +4.782  & 2.3   \\
4 & 0.049 & -5.547 & 9.1   \\
5 & -0.0098 & +5.901  & 43 \\
6 & -0.0151 & -6.094 & 27 \\
7 & -0.00754 & +6.210  & 53  \\
8 & -0.00235 & -6.285 & 168 \\
9 & -0.00039 & +6.337  & 994 \\
\hline
\end{tabular}
\end{table}

\begin{figure}[!htbp]
	\centering
		\includegraphics[width=1\textwidth]{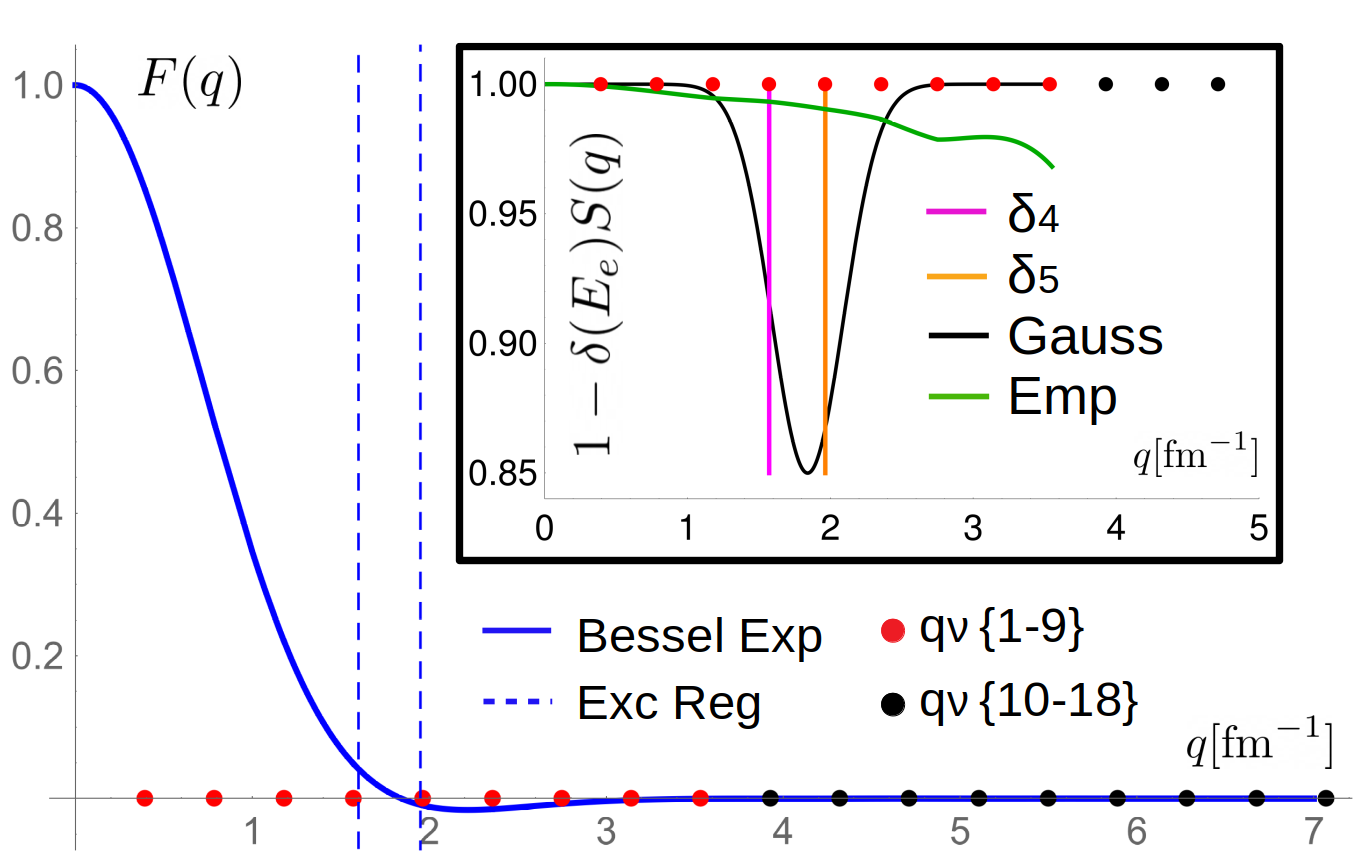}
	\caption{\label{fig:AppPlot}$^{12}$C form factor expanded in the Bessel functions formalism using Offermann \cite{Offermann:1991ft} coefficients without dispersive corrections. The circles in the $q$ axis shows the special values of momentum transfer for the first 9 (red) from experimental data and the second 9 (black) from the extrapolation suggested in \cite{Dreher1974219}. The dashed blue lines encloses the region of the data excluded from the analysis in \cite{Offermann:1991ft}. The inset plot shows the three test forms for $S(q)$ {\color{black}in addition to the empirical perturbation obtained directly from the data by third degree spline interpolation. The curves in the inset plot are the ones needed to obtain the corrected $F_\text{ch}^{stat}$ from the observed $F_\text{ch}^{disp}$ values.} }
	
\end{figure}

As previously stated in the main discussion, we assume in the calculation of $\delta y_i$ that we can separate the effects of the dispersive corrections on the form factor values as (Eq.~\eqref{eq: dispF}): $F_{disp}(q) =F(q)_{stat} [ 1 + \frac{1}{2}\delta (E_e)S(q) ]$ where $\delta(E_e)$ controls the overall strength of the perturbation and $S(q)$ controls the impact this change would have on different $q$ values. Table~\ref{ResultsTable} in the main body shows the results for three different test perturbations $S(q)$, {\color{black} in addition to an empirical one obtained from comparing columns 2 and 3 of Table X in \cite{Offermann:1991ft}}, for the central values of the form factor. {\color{black} For the test perturbations, the central values of the form factor} were modified assuming a constant high value of $\delta (E_e) = 30\%$, so that our analysis could serve as an upper bound. 

The three test forms for $S(q)$ consists of $\delta_4$, $\delta_5$ and Gaussian. The first two represent an {\color{black} up}-shift of $15\%$ on the value of $F(q_\nu)$ for $\nu=4$ and $\nu=5$ alone respectively, while the Gaussian represents a Gaussian {\color{black} up}-shift of amplitude $15\%$ at its peak, centered at the diffraction minimum $q=1.84$ fm$^{-1}$ and with a standard deviation of $0.25$ fm$^{-1}$. The functional forms of the three $S(q)$ are shown in the inset of Fig.~\ref{fig:AppPlot} {\color{black} as well as the empirical perturbation}, while the outset plot shows the Bessel expanded form factor and the special values of the momentum transfer $q_\nu$. 

In all three {\color{black} test} cases for $S(q)$ the change on the radius did not exceed $2\%$, which is still a substantial increase compared to Offermann result \cite{Offermann:1991ft} of a $0.28\%$ increase. {\color{black} The empirical perturbation showed a change of $0.25\%$, consistent with the reported result \cite{Offermann:1991ft}.} This contrast suggests that our overall strength $\delta (E_e)=30\%$ was too large and could imply that for the data range in Offermann work \cite{Offermann:1991ft} $\delta(E_e)S(q)\ll 30\%$, {\color{black} as can be inferred by the small size of the empirical perturbation}. 

{\color{black} This empirical perturbation was only calculated at the special values $q_\nu$ and interpolated using a third degree spline and therefore, is not discarded that it's strength can reach a peak of ~$30\%$ in the excluded region around the diffraction minimum $1.6<q<1.95$ fm $^{-1}$. Indeed, the authors excluded this data to perform their analysis and avoid as much as possible the dispersive effects. }

\end{appendix}

\bibliographystyle{epj}
\bibliography{disp_gueye_rev}

\newpage
\onecolumn
\noindent
\textbf{The Jefferson Lab Hall A Collaboration} \\
\author{
		 P.~Gu\`eye\inst{1}  \and
		A.~A.~Kabir\inst{2} \and \fnmsep \inst{3} \and
		P.~Giuliani\inst{4} \and
		J.~Glister\inst{5} \and \fnmsep \inst{6} \fnmsep \inst{7} \and
		B.~W.~Lee\inst{8} \and
		R.~Gilman\inst{9} \and
		D.~W.~Higinbotham\inst{10} \and
		E.~Piasetzky\inst{11} \and
		G.~Ron\inst{11} \and
		A.~J.~Sarty\inst{5} \and
		S.~Strauch\inst{12} \and
		A.~Adeyemi\inst{3} \and
		K.~Allada\inst{13} \and
		W.~Armstrong\inst{15} \and
		J.~Arrington\inst{15} \and
		H.~Arena\"{o}vel\inst{16} \and
		A.~Beck\inst{17} \and
		F.~Benmokhtar\inst{18} \and
		B.~L.~Berman\inst{19} \and
		W.~Boeglin\inst{20} \and
		E.~Brash\inst{21} \and
		A.~Camsonne\inst{10} \and
		J.~Calarco\inst{22} \and
		J.~P.~Chen\inst{10} \and
		S.~Choi\inst{8} \and
		E.~Chudakov\inst{10} \and
		L.~Coman\inst{23} \and
		B.~Craver\inst{23} \and
		F.~Cusanno\inst{24} \and
		J.~Dumas\inst{9} \and
		C.~Dutta\inst{13} \and
		R.~Feuerbach\inst{10} \and
		A.~Freyberger\inst{10} \and
		S.~Frullani\inst{24} \and
		F.~Garibaldi\inst{24} \and
		J.-O.~Hansen\inst{10} \and
		T.~Holmstrom\inst{25} \and
		C.~E.~Hyde\inst{26} \and \fnmsep \inst{27} \and
		H.~Ibrahim\inst{28} \and
		Y.~Ilieva\inst{19} \and
		X.~Jiang\inst{9} \and
		M.~K.~Jones\inst{10} \and
		A.~T.~Katramatou\inst{4} \and
		A.~Kelleher\inst{29} \and
		E.~Khrosinkova\inst{4} \and
		E.~Kuchina\inst{9} \and
		G.~Kumbartzki\inst{9} \and
		J.~J.~LeRose\inst{10} \and
		R.~Lindgren\inst{24} \and
		P.~Markowitz\inst{20} \and
		S.~May-Tal~Beck\inst{17} \and
		E.~McCullough\inst{5} \and \fnmsep \inst{30} \and
		D.~Meekins\inst{10} \and
		M.~Meziane\inst{30} \and
		Z.-E.~Meziani\inst{15} \and
		R.~Michaels\inst{10} \and
		B.~Moffit\inst{30} \and
		B.~E.~Norum\inst{23} \and
		G.~G.~Petratos\inst{4} \and
		Y.~Oh\inst{8} \and
		M.~Olson\inst{31} \and
		M.~Paolone\inst{12} \and
		K.~Paschke\inst{23} \and
		C.~F.~Perdrisat\inst{30} \and
		M.~Potokar\inst{32} \and
		R.~Pomatsalyuk\inst{10} \fnmsep \inst{31} \and
		I.~Pomerantz\inst{11} \and
		A.~Puckett\inst{34} \and
		V.~Punjabi\inst{35} \and
		X.~Qian\inst{36} \and
		Y.~Qiang\inst{35} \and
		R.~D.~Ransome\inst{9} \and
		M.~Reyhan\inst{9} \and
		J.~Roche\inst{37} \and
		Y.~Rousseau\inst{9} \and
		B.~Sawatzky\inst{10} \and 
		E.~Schulte\inst{8} \and
		M.~Schwamb\inst{16} \and
		M.~Shabestari\inst{23} \and
		A.~Shahinyan\inst{38} \and
		R.~Shneor\inst{11} \and
		S.~\v{S}irca\inst{39} \and
		K.~Slifer\inst{23} \and
		P.~Solvignon\inst{15} \and
		J.~Song\inst{8} \and
		R.~Sparks\inst{10} \and
		R.~Subedi\inst{3} \and
		G.~M.~Urciuoli\inst{24} \and
		K.~Wang\inst{23} \and
		B.~Wojtsekhowski\inst{10} \and
		X.~Yan\inst{8} \and
		H.~Yao\inst{14} \and
		X.~Zhan\inst{34} \and
		X.~Zhu\inst{15} \fnmsep \inst{36}
	   }
\\\\
$^1$			National Superconducting Cyclotron Laboratory, Michigan State University, East Lansing, MI 48824 USA\\
$^2$			Hampton University, Hampton, VA 23668 USA\\ 
$^3$			Kent State University, Kent, Ohio 44242, USA\\ 
$^4$			Florida State University, Tallahassee, FL 32306, USA\\ 
$^5$			Saint Mary's University, Halifax, Nova Scotia B3H 3C3, Canada\\ 
$^6$			Dalhousie University, Halifax, Nova Scotia B3H 3J5, Canada\\ 
$^7$			Weizmann Institute of Science, Rehovot 76100, Israel\\ 
$^8$			Seoul National University, Seoul 151-747, Korea\\ 
$^9$			Rutgers, The State University of New Jersey, Piscataway, New Jersey 08855, USA\\ 
$^{10}$		Thomas Jefferson National Accelerator Facility, Newport News, Virginia 23606, USA\\ 
$^{11}$		Tel Aviv University, Tel Aviv 69978, Israel\\ 
$^{12}$		University of South Carolina, Columbia, South Carolina 29208, USA\\ 
$^{13}$		University of Kentucky, Lexington, Kentucky 40506, USA\\ 
$^{14}$		Temple University, Philadelphia, Pennsylvania 19122, USA\\ 
$^{15}$		Argonne National Laboratory, Argonne, Illinois 60439, USA\\ 
$^{16}$		Institut f\"{u}r Kernphysik, Johannes Gutenberg-Universit\"{a}t, D-55099 Mainz, Germany\\ 
$^{17}$		NRCN, P.O. Box 9001, Beer-Sheva 84190, Israel\\ 
$^{18}$		Duquesne University, Pittsburgh, PA 15282 USA\\ 
$^{19}$		George Washington University, Washington, D.C. 20052, USA\\ 
$^{20}$		Florida International University, Miami, Florida 33199, USA\\ 
$^{21}$		Christopher Newport University, Newport News, Virginia 23606, USA\\ 
$^{22}$		University of New Hampshire, Durham, New Hampshire 03824, USA\\ 
$^{23}$		University of Virginia, Charlottesville, Virginia 22094, USA\\ 
$^{24}$		INFN, Sezione Sanit\'{a} and Istituto Superiore di Sanit\'{a}, Laboratorio di Fisica, I-00161 Rome, Italy\\ 
$^{25}$		Longwood University, Farmville, Virginia, 23909, USA\\ 
$^{26}$		Old Dominion University, Norfolk, Virginia 23508, USA\\ 
$^{27}$		Universit\'{e} Blaise Pascal / CNRS-IN2P3, F-63177 Aubi\`{e}re, France\\ 
$^{28}$		Cairo University, Giza 12613, Egypt\\ 
$^{29}$		University of Western Ontario, London, Ontario N6A 3K7, Canada\\ 
$^{30}$		College of William and Mary, Williamsburg, Virginia 23187, USA\\ 
$^{31}$		Saint Norbert College, Greenbay, Wisconsin 54115, USA\\ 
$^{32}$		Jo\v{z}ef Stefan Institute, 1000 Ljubljana, Slovenia\\ 
$^{33}$		NSC Kharkov Institute of Physics and Technology, Kharkov 61108, Ukraine\\ 
$^{34}$		Massachusetts Institute of Technology, Cambridge, Massachusetts 02139, USA\\ 
$^{35}$		Norfolk State University, Norfolk, Virginia 23504, USA\\ 
$^{36}$		Duke University, Durham, North Carolina 27708, USA\\ 
$^{37}$		Ohio University, Athens, Ohio 45701, USA\\ 
$^{38}$		Yerevan Physics Institute, Yerevan 375036, Armenia\\ 
$^{39}$		Dept. of Physics, University of Ljubljana, 1000 Ljubljana, Slovenia

\maketitle

\end{document}